\documentstyle[12pt]{article}
\begin{document}

\begin{titlepage}

\begin{flushright}
KEK CP-030\\
INP MSU Preprint-95-16/380\\
%hep-ph/95?????\\
14 June 1995
\end{flushright}

\begin{center}
{\large\bf\boldmath  Probing the $H^3$ vertex
   in $e^+e^-$, $\gamma e$ and $\gamma\gamma$ collisions \\
      for light and intermediate Higgs bosons}
\end{center}

\bigskip
\begin{center}
{ V.A.~Ilyin\footnote{ilyin@theory.npi.msu.su},
  A.E.~Pukhov\footnote{pukhov@theory.npi.msu.su} \\
  {\small \it Institute of Nuclear Physics, Moscow State
      University, 119899 Moscow, RUSSIA}

\vskip 1cm
  Y.~Kurihara\footnote{kurihara@kekvax.kek.jp},
  Y.~Shimizu\footnote{shimiz@minami.kek.jp}\\
  {\small \it National Laboratory for High Energy Physics (KEK),
       Tsukuba, Ibaraki 305, JAPAN}

\vskip 1cm
  T.~Kaneko\footnote{kaneko@minami.kek.jp, On leave of absence from
     Meiji-Gakuin University, Kamikurata, Totsuka 244, JAPAN.}\\
  {\small \it Laboratoire d'Annecy-le-Vieux de Physique des Particules} \\
  {\small \it (LAPP), B.P. 110, F-74941 Annecy-le-Vieux Cedex, FRANCE}
}
\end{center}

\vfill

\begin{abstract}
We have studied double Higgs production at future linear colliders while
paying special attention to the option of high-energy and high-luminosity
photon beams.  The main purpose was to examine the feasibility of $e^+e^-$,
$\gamma e$ and $\gamma\gamma$ colliders in order to probe the anomalous
triple Higgs coupling, which is crucial for understanding the Standard Model.
We considered mainly the cases of light and intermediate Higgs bosons.
Double Higgs production is almost background free, except in the $M_H\sim
M_Z$ mass range, which is discussed separately. It is shown that for a light
Higgs boson the $H^3$ coupling can be measured even at $e^+e^-$ collider at
500 GeV. For a intermediate Higgs boson a collider in the TeV region is
suitable for such an investigation.  We have estimated the bounds on the
anomalous $H^3$ coupling, which can be experimentally established using
future linear colliders.
\end{abstract}

%\vfill
%\begin{center}
%{(To be submitted to Nucl. Phys. B)}
%\end{center}
\end{titlepage}
%========================================================
\section {Introduction}

One of the most important problems after the Higgs boson is discovered in the
future will be to study its self-interaction.  It is necessary to clarify the
nature of the spontaneous breaking of the gauge symmetry which provides
nonzero masses of intermediate bosons and fermions.  In the Standard Model
(SM) one scalar Higgs doublet field is introduced,
$$ \Phi\;=\;\frac{1}{\sqrt{2}} \left( \begin{array}{c}
                                          H+\upsilon+i\phi_3 \\
                                          -\phi_2-i\phi_1
                                      \end{array}  \right).$$
Here, $H$ is the physical scalar boson (the Higgs boson, itself) and the
$\phi_i$'s are unphysical Goldstone fields corresponding to pure gauge
degrees of freedom. The Higgs potential is $SU(2)$ invariant,
\begin{equation}
 V(\Phi^\ast\Phi)\;=\;\lambda (\Phi^\ast\Phi-\frac{1}{2} \upsilon^2)^2,
                      \qquad \upsilon=\frac{2M_W\sin{\theta_W}}{e}.
                                                         \label{eq:V2H}
\end{equation}
Here, $e=\sqrt{4\pi\alpha}$ is the electric charge, $M_W$ the mass of $W$
boson and $\theta_W$ the Weinberg mixing angle. The value of the vacuum
expectation, $\upsilon\approx 250$ GeV, is fixed by the parameters of the
intermediate bosons obtained from the experiments.  The coupling constant
($\lambda$) is a free parameter in SM and is associated with the Higgs mass,
$$ \lambda\;=\;\frac{\pi\alpha}{4}
    \frac{M_H^2}{\sin^2{\theta_W} M_W^2}. $$
The experimental bound on the Higgs mass is now $M_H>64.5$ GeV
\cite{LEP-MH}. The region $M_H<90$ GeV ({\it light} Higgs) will be
explored by LEP200 experiments while the mass range up to 400 GeV ({\it
intermediate} Higgs, $M_H<2M_Z$, and heavier) can be scanned at future
linear colliders with $\sqrt{s}=500$ GeV (discussed intensively these
days \cite{H-LC500}).

In SUSY extensions of SM (see, for example, \cite{HHG,H-MSSM} and references
therein) several scalar particles are predicted to have the lightest mass,
less than 200 GeV. The hypothesis of grand unification theories also requires
such light Higgs boson in order to provide the experimental value of
$\sin\theta_W$.  Experiments at LEP200 and future linear colliders will
crucially expose these intriguing theoretical constructions. Hence, in the
situation where only one scalar boson is discovered, a search for evidence of
a nonminimal Higgs self-interaction becomes an actual problem. Such evidence
could be deviations of $H^3$ and $H^4$ couplings from their SM values and
contributions of higher order vertices ($H^n$, $n>4$).  Unfortunately, direct
experimental probing of the vertices of $H^4$ and higher order is impossible,
even at the presently discussed colliders, because of cross sections that are
too small. A measurement of the triple Higgs coupling is thus the only
possibility to confirm the SM structure and to select new theories.

The $H^3$ vertex also contributes to single $H$ production, which is
discussed as discovery reactions.  However, in these processes the
interaction contains {\it Higgs-light fermion} vertex, which has negligibly
small coupling constant. So the $H^3$ vertex contribution to these reactions
is very small.  Furthermore, single $H$ production reactions of higher orders
($\alpha^n$, $n\ge 5$) have very small cross sections.  As a result, {\it the
processes of order $\alpha^3$ and $\alpha^4$ with double Higgs production}
are practically the only way to investigate a Higgs self-interaction.

%%%%%%%%%%%% PLC %%%%%%%%%%%%%%%%%%%%%%%%%%%%%%%%%%%%%%%%%%%%

For this purpose, $e \gamma$ and $\gamma \gamma$ colliders ({\it Photon
Linear Colliders --} PLC), which are based on the idea of Compton
backscattering of laser photons against the electron beam
\cite{e2gam,saldin}, are useful tools as well as $e^+e^-$ colliders are.
The energy distribution of the backscattered photons generally has a wide
spectrum.  If it is possible, however, to shift the interaction point far
enough from the conversion point, the low-energy photons will exit the
interaction area due to relatively large escape angles. The photon beam can
be made practically monochromatic (peaked at a point close to the energy of a
basic electron beam, $E_\gamma^{max}\sim 0.8 E_e$) in this way.  Moreover,
when the polarization of laser photons and that of electrons are opposite,
the spectrum of backscattered photons will be most monochromatic.  Our
analysis is based on the cross sections for a monochromatic beam.  It is
expected that the luminosity of PLC could be on the same order as that for an
$e^+e^-$ collider \cite{fn1}, which is expected to be up to
$100\,\mbox{fb}^{-1}/\mbox{\it year}$.  We also estimate decreasing cross
sections resulting from a convolution with the whole energy spectrum of the
backscattered photon (we used formulas given in \cite{e2gam}).

Electron beams can be highly polarized at future linear colliders
\cite{ee-pol}. A polarized positron beam has also been proposed
\cite{polpos}, though there are still technical difficulties.
Since the polarized positron beam is important, anyway, we also mention the
polarization of the positron beam.  A high rate of circular polarization of
photon beams will also be available. For $e^+e^-$ and $\gamma e$ processes,
the total cross sections crucially depend on electron and positron chirality.
The effective luminosity can be enhanced by using polarized beams.  This
enhancement could be important because much higher statistics are needed for
some physically interesting processes, such as the double Higgs production.
Keeping this circumstance in mind, we note that in $\gamma$ processes, which
we consider here, there is no significant dependence on the polarization of
photons.  However, for $e^+e^-$ and $\gamma e$ processes, the cross sections
in the case of polarized electrons gives just twice (or more when the
positron beam is also polarized) that of the unpolarized one.  We then
basically estimate unpolarized results, and some comments about a
polarization dependence of the cross sections are given, if necessary.

The paper is organized as follows. In section \ref{anomalH}
parametrization of an anomalous Higgs potential is introduced. In section
\ref{HH-prod} we present numerical results for processes and some
comments on signals produced along with double Higgs production.  Those in
$e^+e^-$ collisions have been analyzed in \cite{ee-GSR79}-\cite{ee-BH90}. In
the present work we confirm their numerical results in $e^+e^-$ collisions
(Sec.\ref{ee-HH}) and give new results for $\gamma e$ (Sec.\ref{ea-HH}) and
$\gamma\gamma$ (Sec.\ref{aa-HH}) collisions. We pay special attention to the
case $M_H \approx M_Z$, including a background analysis concerning the
discussed processes in section \ref{MH-MZ}.  In section \ref{delta-analysis}
we give an analysis of the dependence of the cross sections on anomalous
$H^3$ coupling for considered processes in $e^+e^-$, $\gamma e$ and
$\gamma\gamma$ collisions.

%===========================================================
\section{Anomalous Higgs potential \label{anomalH}}

To estimate the contribution of the triple Higgs vertex we have to change the
corresponding constant $\lambda$ in the Higgs potential (\ref{eq:V2H}) while
keeping the $SU(2)$ invariance as well as the value of the vacuum
expectation. We thus add to the SM potential the following monomials
\cite{fn2}:
$$
V_n(\Phi^\ast\Phi)\;\equiv\;\frac{\lambda_n}{n!} (2 \Phi^\ast\Phi-
           \upsilon^2)^n, \qquad n=3,4,\ldots $$
Although many new vertices will appear, for those processes of order
$\alpha^3$ and $\alpha^4$ only some of them can contribute. They are
\begin{equation}
 V_3^{(3)+(4)}\;=\;\frac{\lambda_3}{6}
          (8\upsilon^3 H^3 + 12 \upsilon^2 H^4
 + 12\upsilon^2 H^2 \phi^2_3 + 24 \upsilon^2 H^2 \omega^+\omega^-),
                                                  \label{eq:3H34}
\end{equation}
\begin{equation}
  V_4^{(3)+(4)}\;=\;\frac{2\lambda_4}{3} \upsilon^4 H^4.
                                                  \label{eq:4H34}
\end{equation}
where $\omega^\pm=(\phi_1\mp i\phi_2)/\sqrt{2}$.

In the unitary gauge, where $\phi_i=0$, this new potential changes only two
SM vertices, $H^3$ and $H^4$, and results in new free parameters,
$\lambda_{3,4}$. In other gauges, for example in renormalizable covariant
gauges, all vertices (\ref{eq:3H34},\ref{eq:4H34}) can contribute, again
with two free parameters.  Note that the constant at the $H^4$ vertex stays
a free parameter for $O(\alpha^3)$ and $O(\alpha^4)$ processes.
Unfortunately $\lambda_4$ is out of the experimental study due to small
cross sections for possible processes and only $\lambda_3$ coupling can be
seen.

We now introduce the following dimensionless parameter:
$$
\delta\;\equiv\;\frac{8\upsilon^4}{3 M^2} \lambda_3 .
$$
The value $\delta=-1$ corresponds to the vanishing $H^3$ vertex.

It is clear that the cross sections are of quadratic form in $\delta$,
$$
\sigma(\delta) = \kappa (\delta -\delta_0)^2 + \sigma(\delta_0).
$$
Here, $\delta_0$ corresponds to the minimum of this function.
To determine the function $\sigma(\delta)$ we calculate three
points at $\delta=\pm 1,0$, where $\delta=0$ implies SM.

Since the cross section is a quadratic function of $\delta$, any analysis to
derive an anomalous coupling would be slightly complicated. We now consider
a statistical analysis when the experiment does not show any deviation from
SM at the 95\% CL for $H^3$ coupling. This means that $|N(\delta) - N(0)| <
1.96\sqrt{N(0)}$, where $N(\delta)={\cal L}\sigma(\delta)$ is the number of
detected events and ${\cal L}$ is the integrated luminosity. Two variants
are possible:

\vskip 0.2cm
$(A)$ if $\delta_0^2>D^2$, then
\begin{equation}
\delta_0-\sqrt{\delta_0^2+D^2} <\delta< \delta_0-\sqrt{\delta_0^2-D^2}
  \qquad or \qquad
\delta_0+\sqrt{\delta_0^2-D^2} <\delta< \delta_0+\sqrt{\delta_0^2+D^2};
                                    \label{eq:Abounds}
\end{equation}

\vskip 0.2cm
$(B)$ if $\delta_0^2<D^2$, then
\begin{equation}
\delta_0-\sqrt{\delta_0^2+D^2} <\delta< \delta_0+\sqrt{\delta_0^2+D^2}.
                                    \label{eq:Bbounds}
\end{equation}
Here,
$$
\qquad \qquad D^2 = \frac{1.96 \sqrt{\sigma(0)}}{\kappa \sqrt{\cal L}}.$$
Variant $(A)$ implies that two values of $\delta$ correspond to the
measured cross section (within experimental errors). Furthermore, these two
values are separated by fixed interval which does not depend on the
experimental errors. This discrete uncertainty takes place even if the
number of measured events is sufficient to the level predicted by SM --
some {\it shadow} interval will show up! When the luminosity is small,
variant $(B)$ is realised. However, with increasing the integrated
luminosity a discrete uncertainty appears at some critical integrated
luminosity. It depends on only the Higgs mass, and equals
$$
\hat{\cal L} =\left(\frac{1.96}{\delta_0^2}\right)^2
               \frac{\sigma(0)}{\kappa^2}.$$

To characterize the dependence on $\delta$ we use two parameters which we
denote as $\delta^\pm$. In case $(B)$ these parameters are the corresponding
bounds in (\ref{eq:Bbounds}). In case $(A)$ they are the bounds of either of
the intervals in (\ref{eq:Abounds}) that includes the SM point $\delta=0$.

We cannot expect that the statistics for double Higgs-production reactions
could be high. In some cases, therefore, when the number of events is around
seven or less, the distribution of the event probability will be Poisson
rather than Gaussian. Nevertheless, for definiteness we use the formulas
given above.

%=====================================================================
\section{Double Higgs production cross sections\label{HH-prod}}

In this section we present numerical results for some processes with double
Higgs production.  We carried out the calculations using the framework of the
Standard Model. We used CompHEP \cite{CompHEP} and GRACE \cite{GRACE} (see
also \cite{CompHEP-GRACE}) packages for independent calculations of the
matrix elements and cross sections. These packages provide automatic
computation of the cross sections and distributions in the Standard Model as
well as its extensions at the tree level. All processes were estimated,
including a complete set of diagrams. The calculations were made with 1\%
accuracy, and both packages gave consistent results. Numerical results were
obtained with the following values of physical constants: $\alpha=1/128$,
$M_Z=91.178$ GeV, $\sin{\theta_W}=0.474$.

The values of total cross sections are collected in the Table
\ref{tab:cs-ksi}. In Fig.~\ref{fig:tot-s} we show the  energy dependence of
the total cross sections and in Fig.~\ref{fig:tot-MH} the corresponding Higgs
mass dependence.

%+++++++++++++++++++++++++++++++++++++++++++++++++++++++++++++++
  \subsection{\protect\boldmath $e^+e^-$ collisions \label{ee-HH}}

a) {\Large $e^+e^-\to ZHH$}. This process has been investigated in
\cite{ee-GSR79,ee-BHP88}; we also confirmed their numerical results.
Here, Higgs bosons are produced via bremsstrahlung from the $Z$ boson
(see Fig.~\ref{fig:ZHH-diagr}).
The cross section decreases with energy far enough from the threshold.
At some energy the total cross section has its maximum value of
$\sigma^{tot}_{max}$, depending on the Higgs mass. Higgs and $Z$ bosons
escape at large angles and with practically the same distributions.
Fig.~\ref{fig:ee-angle} gives angular distribution for the Higgs boson.
Since fermion chirality is conserved at the {\it Z-fermion} vertex, the
rate of this reaction may increase by practically twice when electrons
and positrons are polarized.

For this reaction there is no competitive background. A light Higgs boson
decays to a $b\bar b$-pair (more than 90\%) and other fermion pairs.  For
$M_H>150$ GeV the Higgs boson decays mainly into $WW$ or $ZZ$. For $M_H<150$
GeV the main branchings are $ZHH\to Zb\bar b b \bar b \to 6\,jets$, and
$ZHH\to ZWWWW\to 10\,jets$ for heavier Higgs bosons. In these cases suitable
cuts on the invariant masses should be introduced around those points
corresponding to $M_H$, $M_Z$ and $M_W$.  We can say that for light Higgs the
observation of more than 5 events per year is plausible at a 500-GeV $e^+e^-$
collider with ${\cal L}=10\,\mbox{fb}^{-1}$/{\it year}. The initial state
radiation would reduce 7\% of the total cross section.  To study a heavier
Higgs boson, an operation at the cross section maximum is needed to obtain
sufficient statistics.

\vspace{0.3cm}
\noindent
b) {\Large $e^+e^-\to {\bar \nu_e} \nu_e HH$}.  This process was
investigated in \cite{ee-BH90}. Our numerical results are in agreement with
this work.  In this reaction diagrams with the $WW$ fusion (see
Fig.~\ref{fig:WWHHdgrm}) give the main contribution.  The total cross section
increases with energy.  In Fig.~\ref{fig:ee-angle} we give an angular
distribution for Higgs bosons which is smooth with two peaks at $\theta_H\sim
10^\circ$ and $170^\circ$.  This reaction has the biggest cross section among
other processes within the discussed Higgs mass range.

The final states are identified by the decay of Higgs bosons. The main
signature for $M_H<150$ GeV is four jets. For larger masses this is four
gauge bosons ($WWWW$, $WWZZ$, $ZZZZ$) with their subsequent decays, whose
invariant mass distribution shows a peak at $M_H$.  The possible background
could be the reaction $e^+e^-\to HH$, which proceeds at the one-loop level.
It seems, however, that its cross section is very small.  In \cite{GH84} it
is estimated to be less than $0.05$ fb at $\sqrt{s}=250$ GeV and $M_H=100$
GeV, and with visible decreasing at high $\sqrt{s}$.

Since only a left-handed electron and a right-handed positron contribute in
this reaction, if the electron beam (both electron and positron beams) would
be polarized, the statistics would be increased by twice (four times,
respectively).  For unpolarized experiments 34 events per year are
produced for $M_H=150$ GeV at $\sqrt{s}=2$ TeV and ${\cal
L}=100\,\mbox{fb}^{-1}$/{\it year}.  The initial state radiation would
reduce about 20\% of the total cross section.

\vspace{0.3cm}
\noindent
c) Other processes with double Higgs production in $e^+e^-$ collisions have
cross sections that are too small.  Below we summarize the results of the
cross section near to their maximum for the lowest Higgs mass, $M=65$ GeV:

\begin{tabular}{ll}
 $e^+e^-\to \bar t tHH$ & $\sigma^{tot}=0.0632\,\mbox{fb}$
        at $\sqrt{s}=800\,\mbox{GeV}$ and $m_{top}=170$ GeV; \\
 $e^+e^-\to W^+W^- HH$ & $\sigma^{tot}=0.0346\,\mbox{fb}$
                           at $\sqrt{s}=700\,\mbox{GeV}$;\\
 $e^+e^-\to ZZHH$ & $\sigma^{tot}=0.0043\,\mbox{fb}$
                           at $\sqrt{s}=610\,\mbox{GeV}$;\\
 $e^+e^-\to ZHHH$ & $\sigma^{tot}=0.807\cdot 10^{-3}\,\mbox{fb}$
                           at $\sqrt{s}=520\,\mbox{GeV}$;\\
 $e^\pm e^-\to e^\pm e^- HH$ & $\sigma^{tot}=0.133\,\mbox{fb}$
                           at $\sqrt{s}=2\,\mbox{TeV}$.
\end{tabular}

Only the last two reactions, where Higgs bosons are produced via $ZZ$ fusion,
are of some interest for light Higgs bosons.  The reaction $ee\to ZHHH$ could
be of special interest because this is the only place where the $H^4$ vertex
contributes via the {\it Higgs - intermediate bosons} vertices.
Unfortunately, the cross section is too small to observe it.

\vspace{0.3cm}
\noindent
d) We found that for all processes with the {\it W-fusion} mechanism the
energy distribution for outgoing particles has a universal behavior, as
represented in Fig.~\ref{fig:e-distr} (not only for the $e^+e^-$ collision,
but also for the corresponding $\gamma e$ and $\gamma \gamma$ collisions).
We see that Higgs bosons with $M_H<300$ GeV are produced with relatively
small energies, peaking at $\sim (M_H+ M_W)$ (or even less). At the same time
the energy of {\it spectator} neutrinos and/or $W$ boson(s) has a high
maximum at $\sim (\sqrt{s}/2 -M_H-M_W)$ and decreases abruptly to the point
$\sim (\sqrt{s}-M_H)/2$.  The remarkable feature is that the point
$\sqrt{s}/4$ separates the energies of the spectators and Higgs bosons at
about the 80\% level for the $W$ boson and the Higgs boson, and at 70\% for
the neutrino.

For the processes proceeding through the bremsstrahlung mechanism the energy
distributions are similar for the Higgs and $Z$ bosons. We point out only
that the distribution for the $Z$ boson is practically constant, while that
for Higgs bosons increases slowly.

  %+++++++++++++++++++++++++++++++++++++++++++++++++++++++++++++++
  \subsection{\protect\boldmath $\gamma e$ collisions \label{ea-HH}}

a) {\Large $\gamma e^- \to\nu_e W^- HH$}. In this reaction Higgs bosons are
produced via $WW$ fusion (Fig.~\ref{fig:WWHHdgrm}), in a way similar to the
$ee\to\nu\nu HH$ case. The cross section increases with the energy. In total
ten Feynman diagrams contribute to the unitary gauge (we do not count
diagrams with {\it Higgs-electron} vertices due to a negligibly small
coupling constant).

In Fig.~\ref{fig:ea-aa-angle} we give the angular distributions for the Higgs
and $W$ bosons. The Higgs distribution is symmetric. One can see that the $W$
boson escapes mainly in directions close to the photon beam with a peak at
$2^\circ-3^\circ$ (for intermediate Higgs).  The energy of the $W$ boson is
high, on the order of $(\sqrt{s}/2 - M_H-M_W)$ (see Fig.~\ref{fig:e-distr}).
Thus, about 90\% of the events would have decay products of $W$ going in a
forward cone of $5^\circ$, and more than 50\% within a cone of $2^\circ$. The
signature with an additional $W$ boson is free from any background. We
conclude that the event selection has to be of two types: 1) with two
additional high energy jets with invariant mass peaking at $M_W$, 2) with a
large missing energy, on the order of $(\sqrt{s}/2-M_H-M_W)$, and missing
transverse momentum ($>10\,\mbox{GeV/c}$). We also note that a number of
events will include Higgs bosons escaping at angles of less than $15^\circ$.

We conclude that about 20 events per year can be observed for $M_H=150$ GeV
at $\sqrt{s_{\gamma e}}=2$ TeV and ${\cal L}=100\,\mbox{fb}^{-1}/year$.  The
convolution with the photon spectrum decreases the cross section by three
times for $M_H<300$ GeV and $\sqrt{s_{ee}}=2$ TeV.  Since only left-handed
electrons contribute in this reaction, the rate increases along with the rate
of the longitudinal polarization of the electron beam.  However, the
dependence on the photon polarization is small, already only 3\% at
$\sqrt{s}\sim 1$ TeV.

\vspace{0.3cm}
\noindent
b) {\Large $\gamma e\to eZHH$}. The total cross section is
decreasing with the energy and Higgs mass. The maximum is
$\sigma^{tot}_{max}=0.009$ fb for $M_H=65$ GeV at $\sqrt{s}=500$ GeV.
This cross section is too small to investigate experimentally.

   %+++++++++++++++++++++++++++++++++++++++++++++++++++++++++++++++
   \subsection{\protect\boldmath $\gamma\gamma$ collisions \label{aa-HH}}

In $\gamma\gamma$ collisions double Higgs production is possible in
several reactions at the tree level, and one process, $\gamma\gamma\to HH$,
at the one-loop level. All of them are on the order of $\alpha^4$.

\vspace{0.3cm}
\noindent
a) {\Large $\gamma\gamma\to WWHH$}. In this reaction at high energies Higgs
bosons are again produced via $WW$ fusion (see Fig.~\ref{fig:WWHHdgrm}). In
total, thirty one Feynman diagrams contribute to the unitary gauge.

The total cross section increases with energy. The angular distribution is
rather flat for Higgs bosons (see Fig.~\ref{fig:ea-aa-angle}), while $W$
bosons escape close to the collision axis, peaking at angles on the order of
$3^\circ - 4^\circ$ (for intermediate Higgs).  The energy distribution for
$W$ bosons has its maximum at $\sqrt{s}/2-M_H-M_W$ (see
Fig.~\ref{fig:e-distr}). Hence, about 90\% of the events would have decay
products of $W$ bosons going into forward or backward cones of $5^\circ$.  If
particle detection is not easy with such small escape angles, the triggering
must include a large missing energy.  It should be noted that the
$\gamma\gamma\to HH$ reaction gives a background comparable to that of the
signal, if only decay products of Higgs bosons are detected, but without
missing energy.  The reaction $\gamma\gamma\to WWHH$ is free from any
background, except for incorrect combinatorial of jets; for $M_H<150$ GeV the
main branching is four $b$-jets (with an invariant mass peak at $M_H$) plus
jets from two additional $W$ bosons, up to four, and/or a large missing
energy.  In total, $8jet$ events can be detected. For $M_H>150$ GeV the
signature includes up to twelve quark jets with the invariant mass peaking at
$M_W$ and $M_H$.

Our conclusion is that for light and intermediate Higgs a visible number of
events would be produced. For example, for $M_H=150$ GeV about 20 events can
be seen per year at PLC with $\sqrt{s_{\gamma\gamma}}=2$ TeV and ${\cal
L}=100\,\mbox{fb}^{-1}/year$, and about 14 events for $M_H=200$ GeV.

The convolution with the photon spectrum decreases the cross section by
$7-12$ times for $M_H=100-200$ GeV.  The dependence on the
photons polarization is only 7\% at $\sqrt{s}\sim 2$ TeV.

\vspace{0.3cm}
\noindent
b) {\Large $\gamma\gamma\to \bar f fHH$}. Here, $f$ is a fermion. Since the
{\it Higgs-fermion} coupling is proportional to the fermion mass, we
calculated only the $t$-quark case to estimate the upper bound for the cross
sections. The total cross section decreases with energy, and has a maximum
value of $0.07$ fb at $M_H=65$ GeV, $\sqrt{s_{\gamma\gamma}}=850$ GeV and
$m_{top}=170$ GeV. Hence, this reaction can have visible statistics for only
light Higgs, and a year integrated luminosity higher than
$100\,\mbox{fb}^{-1}$.

\vspace{0.3cm}
\noindent
c) {\Large $\gamma\gamma\to HH$}.  This reaction proceeds at the one-loop
level. Analytical results for the amplitudes and a detailed numerical
analysis were carried out in \cite{ee-JT94}.  In the range of Higgs masses
discussed here the main features are the following. The cross-section
dependence on the Higgs mass is weak for $M_H<300$ GeV and $\sqrt{s}>1$ TeV.
The total cross section depends on the photon polarizations.  There are two
kinds of polarized cross sections: parallel ($\sigma^{++}=\sigma^{--}$) and
anti-parallel, ($\sigma^{+-}=\sigma^{-+}$). In the parallel case the
polarized cross section decreases very fast with energy, from $\sim 0.6$ fb
at $\sqrt{s_{\gamma\gamma}}=500$ GeV to $\sim 0.02$ fb at
$\sqrt{s_{\gamma\gamma}}=2$ TeV.  At the same time $\sigma^{+-}$ decreases
very slowly, and is on the order of $0.5-0.3$ fb.  We see that the rate of
this reaction is comparable with that of {\it W-fusion} reactions. However,
as shown in section \ref{delta-analysis}, the sensitivity to the anomalous
$H^3$ coupling is several times weaker for light and intermediate Higgs
bosons.

 %+++++++++++++++++++++++++++++++++++++++++++++++++++++++++++++++
  \subsection{\protect\boldmath $M_H = M_Z$ case \label{MH-MZ}}

When the Higgs mass is equal to the $Z$ mass, we cannot separate the Higgs
signal from the $Z$ background by reconstructing the jet-jet mass;
$b$-tagging is useful to separate the Higgs signal for this case. It can be
done by searching a second vertex from $b$-meson decay by a high-resolution
vertex-detector. We assume that 80\% $b$-tagging efficiency for the
light-quark contamination is 0.5\% and a c-quark 35\% \cite{b-tag}.

In the following analysis we require that at least two $b$-quarks are
tagged for separating a double-$H$ signal.  The tagging efficiency for
tagging two $b$-quarks out of four $b$-quarks from double-$H$ production is
97\%.  The $Z \to b {\bar b}$ decay (15\% branching ratio) and the
miss-identification of $Z \to c {\bar c}$ (12\% branching ratio) are taken
into account.  The background from a light-quark decay of $Z$ is
negligible.  We should note that $b $-tagging would remove the
possible background from $W$ bosons due to a small branching ratio of $W
\to cs$. We hereafter neglect the background from $W$ bosons.

To study the $ee\to ZHH$ process we use a hadronic decay mode of the $Z$
boson. A signal is of 6 jets (four $b$-quark jets from two Higgs bosons are
expected).  The background is the $ee\to ZZ{\cal B}$ processes, where
${\cal B}$ stands for $Z$ or $H$.  At a 500-GeV $e^+e^-$ collider these
processes have total cross sections of 1.15 fb and 0.95 fb, respectively,
while the signal process has 0.305 fb. After $b$-tagging one can expect
the ratio to be $S/B\sim 0.56$.

For $ee\to\nu\nu HH$ the background comes from the $ee\to\nu\nu Z{\cal B}$
process.  At $\sqrt{s}=2$ TeV their total cross sections are equal to 33.8 fb
(${\cal B}=Z$) and 6.48 fb (${\cal B}=H$), while the signal process has 0.73
fb.  Another background source also arises from the $ee\to eeZ{\cal B}$
processes with electrons and positrons being hidden in forward-backward
invisible cones.  Their total cross sections are equal to 4.65 fb (${\cal
B}=Z$) and 1.20 fb (${\cal B}=H$).  However, if we veto electrons in the
visible region and apply a missing momentum cut at, for instance, several
tens of GeV, we can reduce this background to less than 1/1000 with keeping
the almost all signal events. Then we neglect these background hereafter.
After the $b$-tagging one can expect the ratio to be $S/B\sim 0.13$.

For $\gamma e\to\nu WHH$ the background is $\gamma e\to\nu WZ{\cal B}$.
Their cross sections were calculated to be $\sigma^{tot}= 31.1$ fb (${\cal
B}=Z$) and $4.49$ fb (${\cal B}=H$), respectively, at $\sqrt{s_{\gamma
e}}=2$ TeV, while the signal process has 0.361 fb.  Here one can also
expect only a very small ratio of $S/B\sim 0.12$ even with $b$-tagging.

For $\gamma\gamma\to WWHH$ the background is given by $\gamma\gamma\to
WWZ{\cal B}$.  The $\gamma\gamma\to WWZZ$ process has
$\sigma^{tot}= 65.6$ fb\footnote{This result
is in good agreement with \cite{Jikia94} where it was calculated for the
first time.} and 6.35 fb for $WWZH$ at
$\sqrt{s_{\gamma\gamma}}=2$ TeV, while the signal process has 0.337 fb.
Thus, $S/B \sim 0.06$ even with $b$-tagging.

Even if we apply a tighter cut for $b$-tagging, we cannot reduce the
background from $Z \to b {\bar b}$ decay.  The best value for the $S/B$
ratio is given by $\sigma_{signal}/ (\sigma^{ZZ}_{BG}*Br(Z \to b {\bar
b})^2 +\sigma^{ZH}_{BG}*Br(Z \to b {\bar b}))$.  The best values are 0.42,
0.26, and 0.14 for $ee \to \nu \nu HH$, $\gamma e \to \nu WHH$, and $\gamma
\gamma \to WHH$, respectively. We need another method to separate $H$ and
$Z$ events by using the angular distributions of their decay products
\cite{KunStir}. However, this method also throws signal events away.
We thus need higher luminosity colliders.

We may conclude that we must assume difficulties to observe
double Higgs signals for $M_H\sim M_Z$ in {\it W-fusion} reactions.
The situation is better for $ee\to ZHH$; one can hope that double
light Higgs events can be detected even at a 500-GeV $e^+e^-$ collider.

  %=====================================================================
  \section{The $\delta$-dependence     \label{delta-analysis}}

In this section we numerically analyze the sensitivity to the anomalous
$H^3$ coupling. The numerical results for the parameters introduced in
section \ref{anomalH} and their expected bounds are summarized in Table
\ref{tab:cs-ksi}. We also show some representative curves in Figs.
\ref{fig:dl-aa} and \ref{fig:cs_dl_mh}.

One can see a large difference between $ee\to ZHH$ and other reactions
associated with different Higgs-boson production mechanisms in {\it
bremsstrahlung} (Fig.~\ref{fig:ZHH-diagr}) and {\it W-fusion}
(Fig.~\ref{fig:WWHHdgrm}), respectively.

In all cases the minimum point ($\delta_0$) eventually moves to the SM point
$\delta=0$ when the Higgs mass increases. We illustrate this in
Fig.~\ref{fig:dl-aa}. However, we observe that $\delta_0$ moves in the
opposite directions in fusion and bremsstrahlung cases. The SM point is
still rather far from $\delta_0$ for $ee\to ZHH$, while in the fusion cases
the SM point reaches $\delta_0$ at $M_H=300$ GeV for $ee\to\nu\nu HH$, at
$M_H=250$ GeV for $\gamma e\to \nu WHH$ and already at $M_H=200$ GeV for
$\gamma\gamma\to WWHH$.

In Table \ref{tab:cs-ksi} the upper and lower bounds of the $\delta$
($\delta^\pm$) are shown for year-integrated luminosity, ${\cal
L}_y=100\,\mbox{fb}^{-1}$.  One can see the change of modes from $(A)$ to
$(B)$ (see the section \ref{anomalH}) at the points $M_H\sim 200$ GeV for
$ee\to ZHH$ and $M_H\sim 110$ GeV for $ee\to \nu\nu HH$.  This critical point
depends on the luminosity; for example, for ${\cal L}_y=10\,\mbox{fb}^{-1}$
in the reaction $ee\to ZHH$ this point is $M_H\sim 70$ GeV.

\vskip 0.3cm
a) {\bf Light Higgs}. In the process $ee\to ZHH$ for $M_H=65$ GeV at
a 500-GeV $e^+e^-$ collider the limitations are $\delta^-\sim -4.0$ and
$\delta^+\sim 1.8$. The effect of discrete uncertainty is seen (see section
\ref{anomalH}) with a {\it shadow} interval $\sim (-10.2,-4.4)$. Since these
two intervals are very close to each other, the real bounds would be about
$-10<\delta < 1.8$. In the $M_H\sim M_Z$ case we have to recall the large
background, $S/B\sim 0.56$ (see section \ref{ee-HH}), and note that an
integrated luminosity of $10\,\mbox{fb}^{-1}/year$ (3 signal events per
year) would not be sufficient for establishing limitations on $\delta$.

At $\sqrt{s}=2$ TeV and ${\cal L}_y=100\,\mbox{fb}^{-1}$ the possible
limitations on $\delta$ for $M_H=65$ GeV are $-0.42 < \delta <0.61$ in $ee\to
\nu\nu HH$. Here, the effect of a discrete uncertainty shows up rather
clearly, and the {\it shadow} interval is $\sim (2.4,3.4)$. We see that the
$ee \to \nu \nu HH$ reaction is better for determining $\delta$, particularly
for the upper bound, ($\delta^+$).  We have to mention that due to poor
statistics and a very small signal-to-background ratio the case $M_H\sim M_Z$
is practically beyond the scope of experimental plausibility to probe the
anomalous $H^3$ coupling at future linear colliders at the TeV energy range
in {\it W-fusion} reactions.

\vskip 0.3cm
b) {\bf Intermediate Higgs}.  First, one can see that a 500-GeV $e^+e^-$
collider has no feasibility for these Higgs masses in all {\it W-fusion}
processes (see Fig.~\ref{fig:tot-s}). However, for a small area, $M_H<100$
GeV, the reaction $ee\to ZHH$ can give some limitations: for $M_H=100$ GeV
and ${\cal L}_y=10\,\mbox{fb}^{-1}$ the bounds are obtained as
$\delta^-=-8.1$ and $\delta^+=2.0$.

For a linac with $\sqrt{s}=2$ TeV and ${\cal L}_y=100\,\mbox{fb}^{-1}$ we
have the following results.  First the lower bound ($\delta^-$) is
practically independent of the reaction type ({\it W-fusion} ones) and the
Higgs mass. It is given by $\delta^-\sim -0.3$.  As for the upper bound
$\delta^+$ the reaction $ee\to\nu\nu WW$ is better for $M_H<110$ GeV where
$\delta^+\sim 0.6$. In this case a discrete uncertainty appears with a {\it
shadow} interval of $\sim (1.3,2.2)$; however this uncertainty can be
resolved with the help of the $\gamma\gamma$ reaction.  For heavier masses
the $\gamma\gamma$ reaction is the best variant to give the limitations at
the $\delta^+ \sim 0.85 - 0.36$ level for $M_H=120 - 200$ GeV.

We also note that if the PLC luminosity can be increased (for example, as
discussed in \cite{Tel90,TelBal}) the $\gamma\gamma$ reaction can set a
strict limit on $\delta$. In the case of ${\cal L}_y^{PLC}\sim
10^4\,\mbox{fb}^{-1}$, although these bounds can be established at the
$\delta^- \sim -0.05$ and $\delta^+ \sim 0.08$ level, they appear together
with a {\it shadow} interval of $\sim (0.2,0.35)$.

\vskip 0.3cm
c) {\bf Heavy Higgs}.  We have found that in the {\it W-fusion} cases the
dependence of the cross sections on the Higgs mass is changed if $\delta$ is
sufficiently large. We show some representative curves for three values of
$H^3$ anomalous coupling in Fig.~\ref{fig:cs_dl_mh}. This effect can be
explained by the competition of two factors. One factor is associated with
decreasing the total phase space volume by large Higgs masses; the other is
associated with increasing the phase volume where the Higgs propagator can be
regarded as being constant. For a larger $\delta$ the second factor becomes
significant. Thanks to this effect anomalous $H^3$ coupling in {\it W-fusion}
processes becomes detectable for rather large masses.  For example, the value
$|\delta|=1$ allows measurements up to $M_H=750$ GeV at a linac with
$\sqrt{s}=2$ TeV in the $ee\to \nu\nu HH$ reaction.  For such high Higgs
masses the unpolarized total cross section is $\sigma^{tot}(\delta=\pm 1) >
0.015$ fb, while the SM cross section is negligible.  In the $\gamma\gamma\to
WWHH$ reaction a measurement of $\delta=1$ is possible up to $M_H\sim 800$
GeV when $\sigma^{tot}(\delta=1)\sim 0.02$ fb.

\vskip 0.3cm
d) {\Large $\gamma\gamma\to HH$}. Finally, we present some basic numbers
for the $\gamma\gamma\to HH$ reaction using the results from
\cite{ee-JT94}. The important point is that anomalous $H^3$ coupling
contributes only to amplitudes with equal photon polarizations. However, for
such polarizations and for $M_H<300$ GeV the cross section is dominated by
diagrams with a $t$-quark loop, more than 90\% for $\sqrt{s}=500$ GeV and
$\sim 100\%$ for $\sqrt{s}>1$ TeV. Consequently, we have here a different
origin for the $\delta$ dependence from tree level reactions.  As a result,
the sensitivity to $\delta$ is several-times weaker in this reaction compared
with that in {\it W-fusion}.  For example, from the figures presented in
\cite{ee-JT94} one can pick out the following cross sections
for $M_H=250$ GeV: $\sigma^{tot}(\delta=0)\sim 0.104$ fb,
$\sigma^{tot}(\delta=-1)\sim 0.112$ fb and $\sigma^{tot}(\delta=1)\sim 0.1$
fb. These results were obtained at $\sqrt{s_{ee}}=2$ TeV along with a
convolution of the photon energy spectrum.  Due to this convolution the cross
section does not show any large change for relatively light Higgs masses
($<300$ GeV).  Hence, the possible limitations are $\delta^-\sim -4.3$ and
$\delta^-\sim 7.3$.  Such a weak sensitivity means that the interaction of
the Higgs boson with the $W$ boson has a stronger dependence on the anomalous
$H^3$ coupling than does an interaction with fermions.  This difference is
certainly associated with the fact that the {\it Higgs-W} interaction is
deeply involved in the mechanism of a spontaneous breaking of the local gauge
invariance (in the case of anomalous coupling also). One can find a
supporting argument in the figures presented in \cite{ee-JT94}. In fact, for
$M_H>500$ GeV and $\sqrt{s}>$ 1 TeV the main contribution to the amplitudes
with equal photon polarizations comes from the $W$-loop diagrams; in this
mass range the sensitivity on $\delta$ is significantly increased.

%========================================
         \section{Conclusions  \label{conclus}}

The 500-GeV $e^+e^-$ collider will have some possibility to observe double
Higgs production and probe anomalous $H^3$ coupling for only light Higgs
boson.  An experimental study is feasible for the $ee\to ZHH$ reaction with
statistics of more than 5 events per year.  The anomalous coupling may be
bounded at the $-10 < \delta < 2$ level.  In this reaction, even for the
$M_H\sim M_Z$ case, a number of events can be separated from the background.

Stronger limitations on the anomalous $H^3$ coupling for light or
intermediate Higgs bosons can be expected at a linac with $\sim 2$ energy TeV
and a year-integrated luminosity of $\sim 100\,\mbox{fb}^{-1}$.  All three
options ($e^+e^-$, $\gamma e$ and $\gamma\gamma$) are plausible for a
experimental studies of reactions induced by the {\it W-fusion} mechanism:
$ee\to\nu\nu HH$, $\gamma e\to \nu WHH$ and $\gamma\gamma\to WWHH$.  For the
light Higgs boson limitations on the order of $-0.4<\delta <0.6$ are possible
in $ee\to\nu\nu HH$ the reaction. For the intermediate Higgs boson in the SM
the rates are more than 30 events per year for $ee\to\nu\nu HH$ and more than
20 events for $\gamma e$ and $\gamma\gamma$. However, the dependence on
anomalous $H^3$ coupling is stronger in $\gamma\gamma\to WWHH$, where the
limitations can be established at the $-0.3<\delta < 0.6$ level for $M_H\sim
100$ GeV and $-0.3<\delta < 0.36$ for $M_H\sim 200$ GeV.  Anomalous
$\delta=\pm 1$ couplings will be measured up to $M_H\sim 700-800$ GeV in
reactions with {\it W-fusion}.

Probing the anomalous $H^3$ coupling for $M_H\sim 1$ TeV and heavier is
possible only in the $\gamma\gamma\to HH$ \cite{ee-JT94} reaction.

The last conclusion is that if the luminosity of the PLC is much higher
compared with that of the basic linac (in $10-100$ times), $\gamma\gamma\to
WWHH$ should give the best limitations at a level of less than $|\delta|
<0.1$.

%========================================================
\section*{Acknowledgements}

This work was supported in part by the Japan Society for the Promotion of
Science. V.A.I. and A.E.P. were supported also by European association INTAS
(project 93-1180) and International Science Foundation (grants M9B000 and
M9B300). We are also indebted to KASUMI Co.  Ltd. and SECOM Co. Ltd. for the
financial support to our collaboration.

V.A.I. expresses his deep gratitude to Minami-Tateya collaboration (KEK)
for hospitality and for providing him with excellent conditions to stay
and work in Japan.

%############### Bibliography  ####################

\clearpage
\newpage

%%############### TABLES CAPTIONS ####################
\section*{Tables captions}
\begin{table}[h]
\caption{Total cross sections and parameters representing
   the dependence on the anomalous $H^3$ coupling.
   Calculations with unpolarized beams.
   For the $\gamma e$ and $\gamma\gamma$ channels the results are obtained
   for monochromatic photon beam(s).
 \label{tab:cs-ksi}
}
\end{table}

\vskip 2cm
%%############### FIGURES CAPTIONS ####################
\section*{Figures captions}
%\unitlength=1mm

\begin{figure}[h]
%############fig:tot-s
%\begin{picture}(160,50)
%\put(0,0){\epsfxsize=7.5cm \leavevmode \epsfbox{cs_s_65.eps} }
%\put(80,0){\epsfxsize=7.5cm \leavevmode \epsfbox{cs_s_150.eps} }
%\end{picture}
\caption{Energy dependence of the unpolarized monochromatic
total cross sections in the SM:
1) $ee\to ZHH$; 2) $ee\to\nu\nu HH$;
3) $\gamma e\to\nu WHH$; 4) $\gamma\gamma\to WWHH$.
\label{fig:tot-s}
}

%############fig:tot-MH
%\begin{picture}(160,50)
%\put(30,0){\epsfxsize=7.5cm \leavevmode \epsfbox{cs_mh.eps} }
%\end{picture}
\caption{
Unpolarized monochromatic total cross section
as a function of the Higgs mass in the SM.
The curves correspond to the same processes as in
Fig.~{\protect\ref{fig:tot-s}}.
For $ee\to ZHH$  $\sigma^{tot}_{max}$ is calculated at the corresponding
energies; these points are represented as a curve. For
other processes the curves are obtained at $\protect \sqrt{s}=2$ TeV.
\label{fig:tot-MH}
}

%############fig:ZHH-diagr
%\mbox{\epsfig{file=zhh.eps}}
\caption{Feynman diagrams for $e^+e^-\to ZHH$ (diagrams with
{\it Higgs-electron} vertices are not represented due to negligible
coupling constant).
\label{fig:ZHH-diagr}
}

%############fig:ee-angle
%\begin{picture}(160,80)
%\put(30,0){\epsfxsize=7.5cm \leavevmode \epsfbox{ee_h.eps} }
%\end{picture}
\caption{Higgs angular distributions in SM: 1) $ee\to ZHH$ at
   $M_H=65$ GeV and $\protect \sqrt{s}=335$ GeV;
   2) $ee\to\nu\nu HH$ at
      $M_H=150$ GeV and $\protect \sqrt{s}=2$ TeV.
       For both distributions 5000 events were generated.
\label{fig:ee-angle}
}

%############fig:WWHHdgrm
%\mbox{\epsfig{file=wwhh.eps}}
\caption{Feynman diagrams for the fusion mechanism $WW\to HH$
for double Higgs production. For the discussed processes virtual incoming
$W$ bosons are suspended to the initial electron (positron) or photon.
\label{fig:WWHHdgrm}
}

%############fig:e-distr
%\begin{picture}(160,50)
%\put(30,0){\epsfxsize=7.5cm \leavevmode \epsfbox{e_distr.eps} }
%\end{picture}
\caption{Energy distribution for {\it out} particles
   for reactions involving the {\it W-fusion} mechanism.
   Here ${\protect x\equiv 2 E/{\protect \sqrt{s}}}$ and
   $x_{max}$ corresponds to the distribution maximum.
   For Higgs bosons  $x_0= 2M_H/{\protect \sqrt{s}}$,
   $x_{max}<2(M_H+M_W)/{\protect \sqrt{s}}$ and
   $x_1\sim 1-2M_W/{\protect \sqrt{s}}$.
   For the spectators, $x_0=0$ (for neutrino) or
   $2M_W/{\protect \sqrt{s}}$ (for $W$ boson),
   $x_{max}\sim 1-2(M_H+M_W)/{\protect \sqrt{s}}$ and
    $x_1\sim 1-M_H/{\protect \sqrt{s}}$.
   This picture is typical for ${\protect \sqrt{s}\sim 2}$ TeV and
     $M_H=100-300$ GeV.
\label{fig:e-distr}
}

%############fig:ea-aa-angle
%\begin{picture}(160,50)
%\put(0,0){\epsfxsize=7.5cm \leavevmode \epsfbox{nwhh150.eps} }
%\put(80,0){\epsfxsize=7.5cm \leavevmode \epsfbox{wwhh150.eps} }
%\end{picture}
\caption{Angular distributions in SM for
  $\gamma e\to \nu WHH$ (left side) and
  $\gamma\gamma\to WWHH$ (right side).
  For all distributions 5000 events were generated.
\label{fig:ea-aa-angle}
}

%############fig:dl-aa
%\begin{picture}(160,50)
%\put(30,0){\epsfxsize=7.5cm \leavevmode \epsfbox{cs_dl_aa.eps} }
%\end{picture}
\caption{Dependence of unpolarized $\sigma^{tot}$ on the $H^3$ anomalous
coupling for $\gamma\gamma\to WWHH$.
\label{fig:dl-aa}
}

%############fig:cs_dl_mh
%\begin{picture}(160,50)
%\put(0,0){\epsfxsize=7.5cm \leavevmode \epsfbox{dl_m_ee.eps} }
%\put(80,0){\epsfxsize=7.5cm \leavevmode \epsfbox{dl_m_aa.eps} }
%\end{picture}
\caption{Unpolarized $\sigma^{tot}$ {\it vs} Higgs mass:
  left side for $ee\to\nu\nu HH$; right side for
  $\gamma\gamma\to WWHH$.
\label{fig:cs_dl_mh}
}

%############
\end{figure}

\clearpage
\newpage

%############  Tables ####################
{\Large\bf Table}

\vskip 2cm
\begin{tabular}{|l|l|lll|l|l|ll|}
\hline
 Process &$M_H$        & \multicolumn{3}{c|}{Cross sections}
& $\delta_0$ & \raisebox{0ex}[3ex][2ex]{$\hat {\cal L}$} &
$\delta^-$ & $\delta^+$ \\
&{\small GeV} & $\delta=0$ (SM) & $\delta=-1$ & $\delta=1$ & &
     {\small $\mbox{fb}^{-1}$} &
   \multicolumn{2}{c|}{\small (${\cal L}=100\,\mbox{fb}^{-1}$)} \\
           \hline \hline
& & \multicolumn{3}{c|}{\small $\sigma^{tot}_{max}$ fb
            (at $\sqrt{s}$ TeV)} & & & &\\
\cline{3-5}
 & 65&  0.61 (0.335)&  0.41  & 0.87    &  -4.2   &     9& -0.73   &  0.62\\
 & $M_Z$&  0.32 (0.44)& 0.20 & 0.49    &  -3.3   &    21& -0.88   &  0.69\\
   $e^+e^-\to $
 &120& 0.20 (0.56) & 0.12    & 0.32    & -2.9    &    36& -1.1    & 0.77 \\
   \hspace{0.5cm} $ZHH$
 &150& 0.14 (0.7)  & 0.079   & 0.23    & -2.5    &    58& -1.3    & 0.82 \\
 &200& 0.094 (0.95)& 0.050   & 0.17    & -2.0    &   109& -4.8    & 0.85 \\
 &250& 0.072 (1.2) & 0.036   & 0.13    & -1.8    &   143& -4.4    & 0.86 \\
 &300& 0.059 (1.5) & 0.029   & 0.12    & -1.6    &   180& -4.1    & 0.85 \\
           \hline \hline
& & \multicolumn{3}{c|}{\small $\sigma^{tot}$ fb
            at $\sqrt{s}=2$ TeV} & & & & \\
\cline{3-5}
 & 65&  1.0    &  1.6    & 0.77    &  1.5    &    42&-0.42    & 0.61    \\
 & $M_Z$& 0.73 &  1.4    & 0.52    &  1.0    &    64&-0.34    & 0.55    \\
           $e^+e^-\to$
 &120& 0.49    &  1.2    & 0.41    & 0.64    &   133&-0.30    &  1.6    \\
    \hspace{0.3cm} $\bar\nu_e\nu HH$
 &150& 0.34    & 0.99    & 0.37    & 0.45    &   262&-0.28    &  1.2    \\
 &200& 0.19    & 0.81    & 0.40    & 0.25    &  1116&-0.27    & 0.77    \\
 &250& 0.11    & 0.67    & 0.41    & 0.15    &  4871&-0.27    & 0.56    \\
 &300& 0.066   & 0.55    & 0.41    & 0.086   & $>10^4$&-0.27    & 0.44  \\
           \hline
 & 65& 0.51    & 0.86    & 0.37    &  1.1    &   100&-0.47    &  2.8    \\
 & $M_Z$& 0.36 & 0.78    & 0.28    & 0.73    &   169&-0.38    &  1.8    \\
           $\gamma e\to$
 &120& 0.26    & 0.71    & 0.27    & 0.47    &   381&-0.33    &  1.3    \\
   \hspace{0.3cm} $\nu_e WHH$
 &150& 0.19    & 0.65    & 0.31    & 0.30    &  1088&-0.32    & 0.91    \\
 &200& 0.11    & 0.58    & 0.38    & 0.13    & $>10^4$&-0.31    & 0.58  \\
 &250& 0.072   & 0.51    & 0.44    & 0.047   & $>10^4$&-0.32    & 0.41  \\
 &300& 0.048   & 0.45    & 0.45    &-0.0012  & $>10^4$&-0.33    & 0.33  \\
           \hline
 & 65&    0.46 & 0.86    & 0.38    & 0.76    &   213&-0.43    &  1.9    \\
 & $M_Z$& 0.34 & 0.86    & 0.36    & 0.47    &   370&-0.34    &  1.3    \\
           $\gamma\gamma\to$
 &120& 0.25    & 0.86    & 0.43    & 0.28    &  1102&-0.30    & 0.85    \\
   \hspace{0.3cm} $WWHH$
 &150& 0.20    & 0.86    & 0.55    & 0.15    &  5817   &-0.29    & 0.59 \\
 &200& 0.13    & 0.84    & 0.76    & 0.028   & $>10^4$ &-0.30    & 0.36 \\
 &250& 0.10    & 0.80    & 0.92    &-0.038   & $>10^4$ &-0.33    & 0.25 \\
 &300& 0.08    & 0.75    & 0.99    &-0.078   & $>10^4$ &-0.35    & 0.20 \\
 &500& 0.033   & 0.36    & 0.61    &-0.14    &  1641   &-0.45    & 0.17 \\
           \hline
\end{tabular}

\vskip 1cm
\centerline{Table 1:}


\begin{thebibliography}{99}

\bibitem {LEP-MH}
    M.~Pohl, {\it Proceedings of 27th Int. Conf. on High Energy Physics}
      (Glasgow, July 20-27, 1994), p.107, ed. P.J.~Bussey and I.G.~Knowls.

\bibitem {H-LC500}
    See, for example, articles in:
    {\it Physics and Experiments with Linear Colliders}, ed. R.~Orava,
          P.~Eerola and M.~Nordberg (World Scientific, Singapore, 1992); \\
    {\it Proc.  Workshop `$e^+ e^-$ collisions at $500GeV$: the physics
          potential'}, ed. P.W.~Zerwas, DESY 92-123A (1992); \\
    `JLC-I', KEK-Report 92-16 (1992).

\bibitem {HHG}
    J.F.~Gunion, H.E.~Haber, G.~Kane and S.~Dawson,
         {\it The Higgs Hunter's Guide} (Addison-Wesley, 1990).

\bibitem {H-MSSM}
    A.~Djouadi, J.~Kalinowski and P.W.~Zerwas,
        in: {\it Proc.  Workshop `$e^+ e^-$ collisions at
               $500GeV$: the physics potential'}, ed. P.W.~Zerwas,
               DESY 92-123A (1992) p.83.

\bibitem {e2gam}
    I.F.~Ginzburg, G.L.~Kotkin, V.G.~Serbo and V.I.~Telnov,
         Pis'ma Zhh. Eksp. Teor. Fiz. {\bf 34} (1981) 514;
         Nucl. Instr. Methods {\bf 205} (1983) 47. \\
    I.F.~Ginzburg et al., Nucl. Instr. Methods {\bf 219} (1984) 5.

\bibitem{saldin}
    E.L.~Saldin et al., DESY 94-243, December 1994.

\bibitem{fn1}
    We are grateful to I.F.Ginzburg for the discussion of this point. The
    corresponding references are \cite{e2gam,Tel90,GKST}. Recent discussion
    has been concluded \cite{aa95} that PLC luminosity can be 1/5 - 1/3 of
    basic linac with practically monochromatic photon beam.  We refer also
    to \cite{Tel90,TelBal} where an idea to increase PLC luminosity
    up to $10^{35}-10^{36}\,\mbox{cm}^{-2}\mbox{s}^{-1}$ is described.

\bibitem{Tel90}
    V.I.~Telnov, Nucl. Instr. Methods {\bf A294} (1990) 72.

\bibitem{GKST}
    I.F.~Ginzburg, Nucl. Phys. B (Proc. Suppl.)
           {\bf 37B} (1994) 303. \\
    I.F.~Ginzburg, G.L.~Kotkin, V.G.~Serbo and V.I.~Telnov,
         Phys. Rep. {\bf C} (to be appeared).

\bibitem{aa95}
    {\it 10th Workshop on Photon-Photon Collisions (Photon'95)}
         (Sheffield, England, April 8-13, 1995).

\bibitem{TelBal}
    V.I.~Telnov, in: {\it Proc. Workshop on Physics and Experiments with
       Linear $e^+e^-$ Colliders}, vol. II,
       ed. F.A.~Harris et al. (World Scientific, Singapore, 1993) p.551;
       Nucl. Instr. Methods {\bf A355} (1995) 3. \\
    V.E.~Balakin and N.A.Solyak, Nucl. Instr. Methods {\bf A355}
        (1995) 142.\\
    V.E.~Balakin and A.A.Sery, Nucl. Instr. Methods {\bf A355} (1995) 157.

\bibitem{ee-pol}
    T.~Maruyama et al., Phys. Rev. Lett. {\bf 66} (1991), 2351, \\
    T.~Omori et al., {\it Proc. of High Energy Accelerator Conference,
       Hamburg 1992}, (World Scientific, Singapore, 1992), Vol-I, p.157, \\
    H.~Aoyagi et al., Phys. Lett. {\bf A167} (1992), 415, \\
    Y.~Kurihara et al., Jpn. J. Appl. Phys. {\bf 34} (1995), 355.

\bibitem{polpos}
    For a helical undulater method, see
        V.E.~Balakin, A.A.~Mikhailichenko, INP 79-85(1979),\\
        K.~Fl\"{o}ttmann, DESY 93-161(1993). \\
    For the other methods, see M.~Chiba et al., in
        {\it Proceedings of the fifth JLC workshop}, ed. Y.~Kurihara.


\bibitem {ee-GSR79}
    G.J.~Gounaris, D.~Schildknecht and F.M~Renard, Phys. Lett. {\bf 83B}
            (1979) 191.

\bibitem {ee-BHP88}
    V.~Barger, T.~Han and R.J.N.~Philips, Phys. Rev. {\bf D38} (1988) 2766.

\bibitem {ee-BH90}
    V.~Barger and T.~Han, Mod. Phys. Lett. {\bf A5} (1990) 667.

\bibitem{fn2}
    An analogous procedure was used in
    \cite{vanderBij86} so as to see the contribution of higher order
    monomials for heavy Higgs ($M_H\sim 1$ TeV and more).

\bibitem {vanderBij86}
    J.J.~van~der~Bij, Nucl. Phys. {\bf B267} (1986) 557.

\bibitem{CompHEP}
    E.E.~Boos et al., Preprint INP MSU-94-36/358, SNUTP 94-116,
          Seoul, 1994; hep-ph/9503280.

\bibitem {GRACE}
    GRACE Manual, v.1.0, KEK Report 92-19, 1993.

\bibitem {CompHEP-GRACE}
    E.E.~Boos et al., Int. J. Mod. Phys. {\bf C5} (1994) 615.

\bibitem {GH84}
    K.J.F~Gaemers and F.~Hoogeveen, Z. Phys. {\bf C26} (1984) 249.

\bibitem {ee-JT94}
    G.~Jikia and A.~Tkabladze, in: {\it Proc.  Workshop `$e^+ e^-$
      collisions at $500GeV$: the physics potential'}, ed. P.W.~Zerwas,
      DESY 93-123C (1993) p.529.

\bibitem {b-tag}
    N.~Brown, Z. Phys. {\bf C49} (1991) 657. \\
    `JLC-I', KEK-Report 92-16 (1992) p.86. \\
    P.~Grosse-Wiesmann, D.~Haidt and J.~Schreiber, in {\it Proc.
       Workshop `$e^+ e^-$ collisions at $500GeV$: the
       physics potential'}, ed. P.W.~Zerwas. DESY 92-123A (1992) p.37.

\bibitem {Jikia94}
    G.~Jikia, IHEP 94-77, hep-ph/9407393, July 1994;
      Nucl. Phys. {\bf B437} (1995) 520. \\
    K.~Cheung, Phys. Rew. {\bf D50} (1994) 4290.

\bibitem {KunStir}
    Z.~Kunszt and W.~Stirling, Phys. Lett. {\bf B242} (1990) 502.

\end{thebibliography}
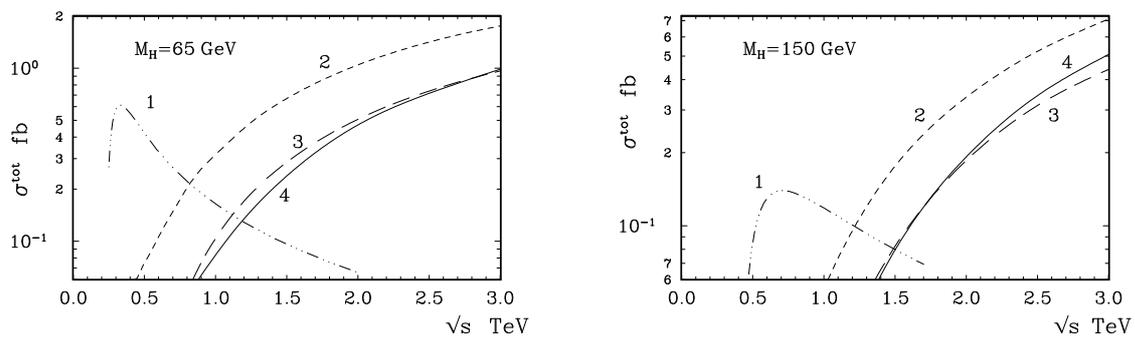
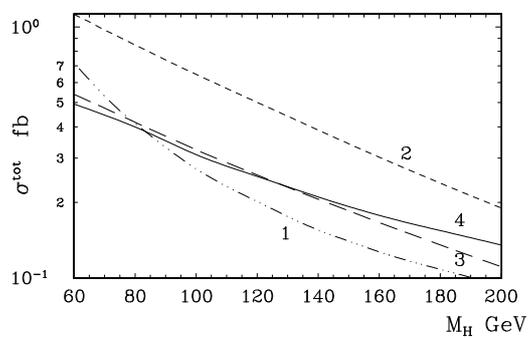
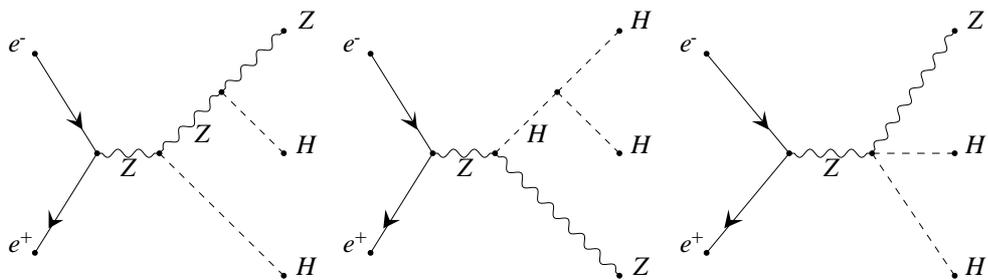
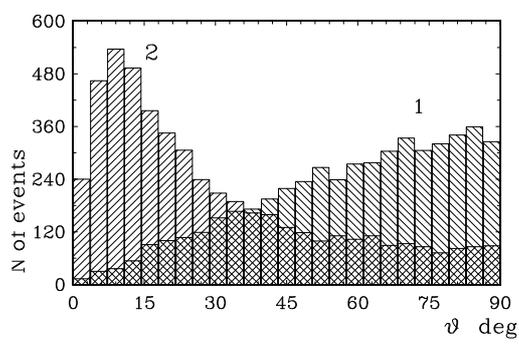
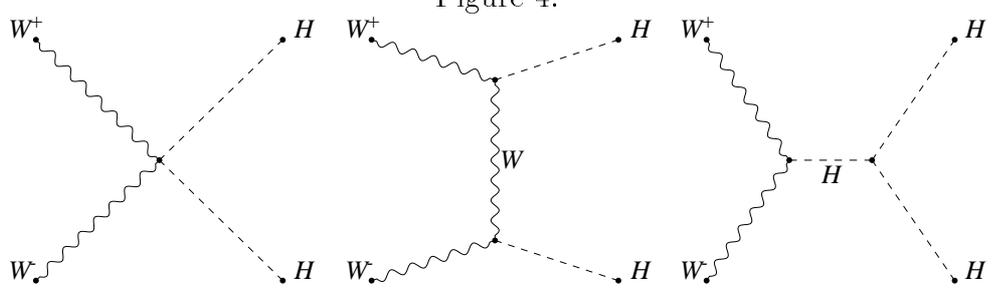
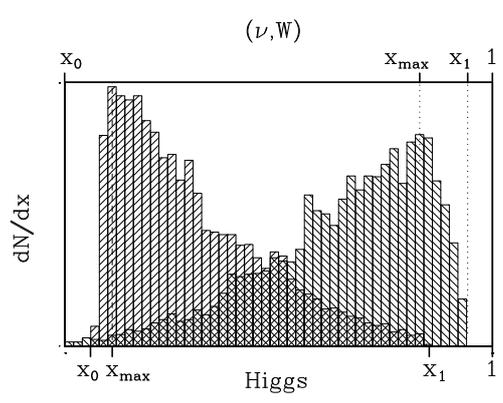
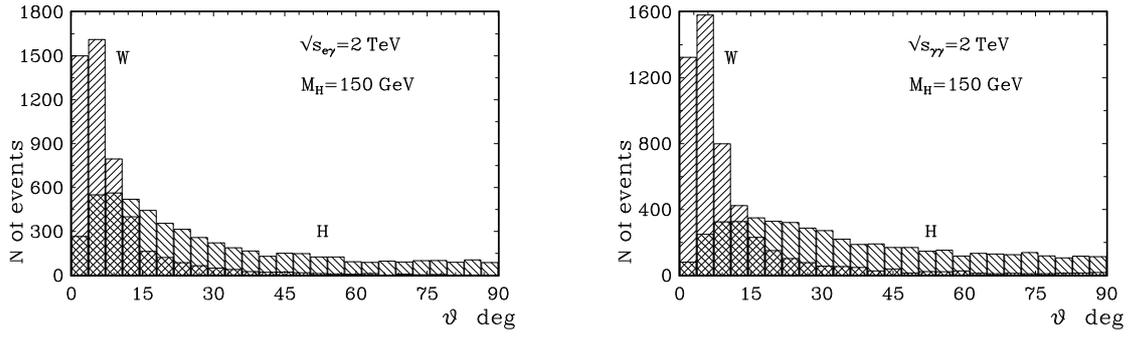
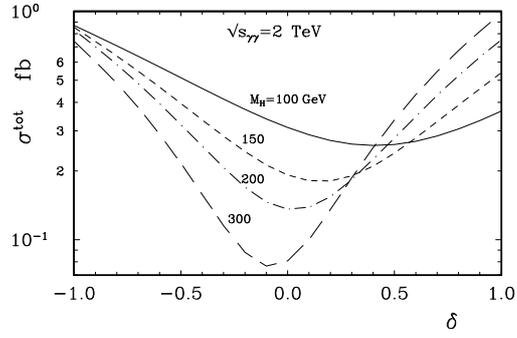
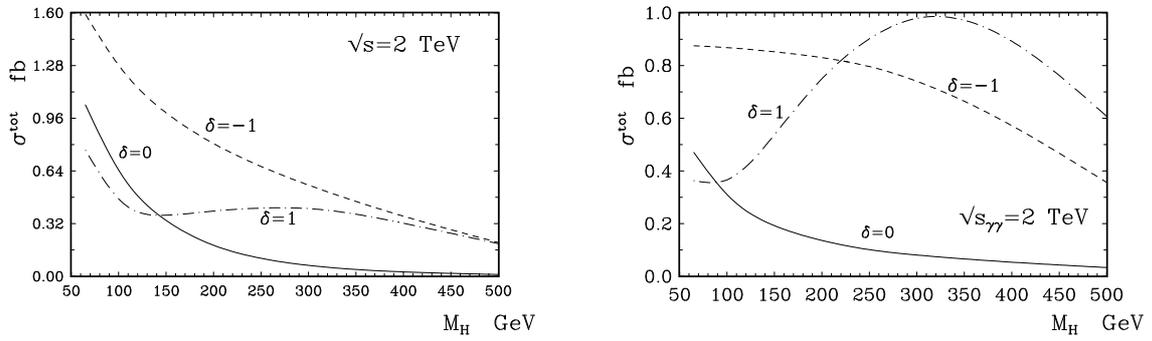
\end{document}